\documentstyle[preprint,prl,aps,epsf]{revtex}
\begin{document}
\draft
\title{
 Quantum Monte Carlo Evidence for d-wave Pairing in the 2D
 Hubbard Model at a van Hove Singularity
}
\author{
T. Husslein$^+$, I. Morgenstern$^+$, D.M. Newns,
P.C. Pattnaik, J.M. Singer$^+$, and H.G. Matuttis$^+$
}
\address{
IBM Thomas J. Watson Research Center \\
P.O. Box 218\\
Yorktown Heights, NY 10598\\
}
\date{\today}
\maketitle
\begin{abstract}
We implement a Quantum Monte Carlo calculation for a repulsive
Hubbard model with nearest and next-nearest neighbor hopping
interactions on clusters up to $12\times 12$.
A parameter region where the Fermi level lies close to the van Hove
singularity at the Saddle Points in the bulk band structure is
investigated.  A pairing tendency in the
$d_{x^2-y^2}$
symmetry channel, but no
other channel, is found.
Estimates of the effective pairing interaction
show that it is close to the value required for a 40 K superconductor.
Finite-size scaling compares with the attractive
Hubbard model.
\end{abstract}
\narrowtext

\vglue 4cm

\footnotesize{$^+$
Permanent Address: Fakult\"at Physik, Universit\"at Regensburg,
D-93053 Regensburg, Germany}
\newpage
The 2D Hubbard model (HM) contains several basic elements of the high
temperature superconductivity problem, and its properties include
nontrivial features (e.g. the Mott transition and associated
antiferromagnetism) which are generic in HiTc materials.
Is the HM also a superconductor, or does the superconductivity originate
from extrinsic interactions or degrees of freedom?
According to calculations~\cite{1},\cite{2}
based on exchange of antiferromagnetic spin fluctuations (AFSF),
the HM is a superconductor with
$d_{x^2-y^2}$
order parameter symmetry (a symmetry consistent with several
recent experimental measurements~\cite{3}-\cite{5}).
But on the contrary Quantum Monte Carlo
(QMC) calculations up to the present~\cite{6}-\cite{9}
have given negative or inconclusive results
regarding superconductivity of the 2D HM.\\

In the present QMC study of a generalized HM (which includes
next-nearest neighbor hopping interactions as well as the usual
nearest-neighbor ones),
we present evidence for a superconducting tendency in the
$d_{x^2-y^2}$ channel in large
(up to $12\times12$) clusters, in qualitative agreement with the
the AFSF calculations.
  The results have been obtained close to the
parameter region
where the energy of the saddle points (SP) in the band structure
lie close to the 'Fermi level' in the cluster energy level structure,
a choice motivated by the hypothesis~\cite{10}-\cite{12}
- 'Van Hove Scenario' -
that, in the continuum limit, superconductivity is enhanced by the
Van Hove density of states peak associated with the saddle points.\\

The SP feature may be incorporated into the Hubbard model within the
metallic regime by introducing a next-nearest neighbor interaction.
This allows the SP to lie at the Fermi level at a doping of, say 15-25\%,
while the insulating point, at which the antiferromagnetic instability
occurs, lies at 0\% doping.  These features are characteristic of real
cuprate materials~\cite{13},\cite{14}.
It is in the former situation (15-25\% doping)
that the model is found to support superconductivity.

The model is specified as follows
\begin{eqnarray}
H  =  - t \sum_{\langle i j \rangle \sigma} c^{+}_{ i \sigma}
c^{}_{j \sigma}  +  t' \sum_{\langle\langle i j \rangle \rangle \sigma}
 c^{+}_{ i \sigma} c^{}_{j \sigma} +
U \sum_{i} n_{i \uparrow} n_{i\downarrow}
\label{eqno(1)}
\end{eqnarray}
In (1), $U$ is the repulsive on-site Coulomb interaction, $t$ is the
nearest-neighbor hopping integral
($\langle ij \rangle$ denotes nearest neighbor interactions),
$t'$ is the next-nearest neighbor hopping integral
($\langle\langle i j \rangle \rangle$ denotes next-nearest
neighbor interactions); $t$ and $t'$ are defined to be positive.\\

The noninteracting band structure of the
$tt'$-Hubbard model (1) has saddle points at
energy $-4t'$, and at $k =
(0,\pi)$ and $(\pi,0)$.
If we take
the hole doping $x$ to be of order $x \approx t'$, then the
saddle points in the noninteracting
band structure lie near $E_F$ (see Fig. 1).
In the absence of $t'$, the required doping would
be zero, making the sample insulating.
The electronic effective
mass below the SP's is heavier than the mass above, the ratio being
$(t+2 t')/(t-2 t')$.  Photoemission data indeed
show a large mass ratio~\cite{13},\cite{14}.
The simulation of the $tt'$-Hubbard model~\ref{eqno(1)} is based on
the
projector Monte Carlo technique, using the Ansatz for the ground
state~\cite{7}
\begin{eqnarray}
\vert \Psi_g \rangle  =  e^{\Theta H} \vert \Psi_0 \rangle; \Theta \to \infty
\label{(2)}
\end{eqnarray}
where $\Theta$ is a projection parameter
and $\vert  \Psi_0 \rangle $ a single determinant taken as the ground
state of the noninteracting band structure of~\ref{eqno(1)}.\\

The exponent in (2) is broken up into Trotter slices on the interval
$\Theta$, and the Hirsch-Hubbard-Stratonovich~\cite{15}
transformation
applied to each slice, turning (2) into a path integral which is
solved on a finite cluster (up to $12\times 12$) using the Metropolis algorithm
(for details see Ref.~\cite{7}).
We checked on the convergence of the algorithm by comparing small clusters
to exact diagonalization and stochastic diagonalization~\cite{preprint} and
found
$\Theta=8$ and $\tau=0.125$ to be sufficient.
 Periodic boundary conditions and
closed shell configurations are always used.
The 'average sign' is usually sufficiently close to unity, that
the simulations are not significantly limited by the fermionic
sign problem. In a typical simulation we averaged over four runs with different
seeds
each run having at least $10^6$MCS. The nonexistence of a fermionic sign
problem in the simulations and the very large number of MCS allowed
us to obtain precise results for the superconducting correlation functions.
The orthogonalization technique originally proposed by Sorella et al.~\cite{26}
helped us to stabilize the algorithm.
We found a stabilization every 8 Trotter-slices to be sufficient to make
the results independent of the number of stabilizations.\\

The tendency towards superconductivity
is signalled~\cite{9},\cite{16} by the
presence of a long-range 'plateau'
in the superconducting correlation 'function'~\cite{9} plotted
as a function of distance.
Such plateaus have been demonstrated~\cite{7} in
models such as the attractive Hubbard model and simplified
models with electron-phonon coupling, where a pairing tendency
is anticipated to occur.
While studying finite systems one has always to extrapolate to
the thermodynamic limit. Due to the unsystematic finite level
structure of the finite Hubbard model this has turned out to be very difficult.
We use two different approaches:
a) we are able to analyse the plateau in terms of the effective
pairing interaction J, and hence deduce a value for $T_c$
in the infinite-sample limit.
b) we find analogous finite size scaling
behaviour for the superconducting correlations for
the repulsive and the attractive Hubbard model. \\

The condition as to whether
the finite cluster of size $L$ is 'superconducting' (correlation
length $\xi > L$), or whether superconducitivity is
suppressed by finite size effects
($\xi < L$), does not enter into these considerations
and is not relevant for this paper. \\

Some constraints on U are a) according to exact diagonalization
calculations the Hubbard model is an insulator at half-filling,
as observed experimentally, if U is greater~\cite{17}
than about $10 t'$
b) the value U=2t has been used in
calculations~\cite{18} of the quasiparticle lifetime broadening
for the $tt'$-Hubbard model
in the Renormalized Propagator Approximation, with results
consistent with experimental data
c) attempts to derive the single-band U formally
from a multi-band model~\cite{19} give a
value of order 6eV, i.e. 6t, in the case t = 1 eV.
We have observed clear evidence for the superconducting tendency
for U in the range $0.5t < U < 3t$;
for larger U-values unnacceptable error bars are obtained. \\

The superconducting correlation function $\chi(R_j)$ for
$d_{x^2-y^2}$symmetry is defined as
\begin{eqnarray}
\label{(3)}
\chi (R_j)&=& \frac{1}{4N_L} \sum_{i}
\sum_{\langle p \rangle , \langle q \rangle}
\langle [ c^{+}_{i\uparrow}
c^+_{i+p\downarrow}
- c^+_{i\downarrow}
c^+_{i+p\uparrow}] \\
&\times& [ c^{}_{i+j+q\downarrow}
c^{}_{i+j\uparrow}
- c^{}_{i+j+q\uparrow}
c^{}_{i+j\downarrow}] \rangle
\sigma^{}_p \sigma^{}_q  -  \chi^{}_0(R^{}_j)\nonumber
\end{eqnarray}
where
\begin{eqnarray}
\chi^{}_0(R^{}_j)&=&\frac{1}{4N_L} \sum_{i,\sigma}
\sum_{\langle p \rangle , \langle q \rangle}
\left\{ \langle c^+_{i\sigma} c^{}_{i+j\sigma} \rangle
\langle c^+_{i+p -\sigma}  c^{}_{i+j+q -\sigma}\rangle \right. \nonumber\\
&+& \left. \langle c^+_{i\sigma} c^{}_{i+j+q \sigma} \rangle
\langle c^+_{i+p -\sigma} c^{}_{i+j -\sigma} \rangle
\right\} \sigma^{}_p \sigma^{}_q \nonumber
\end{eqnarray}
Here, $\langle p \rangle$ implies a sum over nearest neighbors
of $0$, $p=(\pm 1, 0)$ and $(0, \pm 1)$,
for which $\sigma_p$ is respectively 1 and -1, $N_L$
is the number of atoms in the cluster, and $\sigma$ is spin.\\

A typical calculation of the superconducting correlation function (3)
for a ($12\times 12$) cluster is illustrated in
Fig. 2a.
It is seen that, as a function of increasing distance,
the correlation function in this symmetry
approaches a 'plateau'.  It is only in
 $d_{x^2-y^2}$symmetry that we find
this behavior.
The error bars in the points, calculated as the deviation between runs with
different seeds,
 are smaller than their diameter.

A study of the variation of $\chi^{pl}$
with $U$ is shown in Fig. 2b, based on $8\times 8$ clusters.
$\chi^{pl}$ is seen to increase with U, and
continues to increase steadily up to
the largest value of U (U=3t) for which we have adequate
statistics.\\

The main problem with scaling our results is the influence of  finite
shell structure on the observables and the unsystematic behaviour of
these shells with system size.
Therefore the application
of finite size scaling especially for weak $U$
is extremely complex, as has already been observed for the case of the
attractive Hubbard model~\cite{24}.

In Fig. 3 we illustrate a plot of the value of
$\chi(R)$ averaged over the plateau region,
$\chi^{pl}$, versus $t'$ for a
$10\times 10$ cluster with $N_e=74$ electrons.  Each point in
Fig.3 represents a run with several different seeds,
typically 4.  Fig. 3 contains points calculated using two
different sampling algorithms, which are seen to agree quite
well.  The strong variation of $\chi^{pl}$
with $t'$ seen in Fig. 3 is systematic,
and is due to the shell structure of the cluster.
The peaks in $\chi^{pl}$ are found to be associated with
the near-degeneracy of several filled and empty cluster levels,
i.e. a high 'density of states at $E_F$'.
The dips in $\chi^{pl}$ coincide with regions of
maximum gap in the single particle spectrum of the cluster
between the highest occupied and lowest unoccupied
levels.\\

A more quantitative analysis of the effect of shell structure on
pairing is possible by introducing a simple low energy pairing
Hamiltonian~\cite{20}
\begin{eqnarray}
\label{(4)}
H_p = \sum_{k\sigma} \varepsilon_k
n_{k\sigma}  +
J \sum_{k \neq k'}
p^+_{k}p^{}_{-k'}
\eta^{}_{k} \eta^{}_{-k'},
\end{eqnarray}
Here $k=a$ discrete wavevector for the cluster
(periodic boundary conditions assumed) lying within the
cutoff $\vert \varepsilon_k -\mu \vert < \omega_c$,
$p^+_k = c^+_{k\uparrow}
c^+_{-k\downarrow}$, $\eta_k=(\cos(k_x)-cos(k_y))$, and $J$, which is negative,
is a nearest-neighbor
attractive interaction which parametrizes the pairing strength in
the model (4).\\

For the present purpose we exploit the cutoff $\vert \varepsilon_k
- \mu \vert < \omega_c$, to solve (4)
by exact diagonalization for states within $\omega_c$.
Having available the exact system ground state we
can calculate $\chi^{pl}$, which is also illustrated
in Fig. 3 for J=0.055t.  It is seen that the low energy Hamiltonian
(4), with its shell-structure effects, well reproduces the
corresponding structure in the QMC results.
Using the complete eigenvalue and eigenvector spectrum of the
of $H_p$, within the cutoff,
the projector formula for $\chi^{pl}$,
Eq. (2), can be evaluated explicitly for a given $\Theta$;
we find that the value $\Theta= 1/8t$ adequately
reproduces the ground state value of $\chi^{pl}$.\\

In Table I we demonstrate a possible technique for scaling QMC
$\chi^{pl}$ data on different cluster sizes.
The problem is to avoid the pronounced cluster shell-structure effects
(Fig. 3).  The technique is to compare J, obtained
by fitting the Eq. (4)
model to $\chi^{pl}$ at each cluster size, rather than
$\chi^{pl}$, for different sizes, whereby
the shell-structure effects are taken into account.
Values of J deduced from QMC calculations on $n\times n$ clusters
with n from 6 to 12 are given in Table I.  It is seen that
the results appear to be converging on J=-0.15t,
though results on a wider dynamic range of cluster sizes are
needed to be convincing that convergence on the bulk limit has
been achieved. \\

The foregoing method of analysis is extensible to
include the case of a constant Z-factor (Eliashberg notation).
The analysis yields $J/Z^2$.  A constant DOS
$\rho_0$ renormalizes to $Z \rho_0$,
giving $Z/(J \rho_0)$ in the exponent of the
the $T_c$ equation, which is the correct form
and hence does not require any additional correction for Z.
It is therefore valid, to within the constant-Z approximation,
to proceed as if in the BCS case (Z=1).
The largest value of J we have found is in the case
$t'=0.22$, $U=3$, (see inset Fig. 3), for which $J=-0.2t$.
This value of $J$ is close to that ($J=-0.22t$)
required to give a 40 K $T_c$, with $t=1$ eV.\\

In Figure 4 we present a second method to extrapolate the data. We
found in agreement with~\cite{24} that scaling laws of the form $1/N$ are
too simple and can lead to ambigous results.
Instead of incorporating corrections to scaling while applying the scaling
laws~\cite{preprint}
we directly compare the results on $\chi_{d_{x^2-y^2}}^{pl}$ for positive U
with
results on $\chi_{s_{os}}^{pl}$ for negative U.
 In both cases we use the same system sizes, fillings,
and $t'$. Hence the finite size structure is identical. Additionally we use a
value of U=-0.3
to match the size of the plateaus for the positive U ($d_{x^2-y^2}$) case.
Our analysis shows that the attractive and repulsive HM have the same scaling
behaviour.
As the attractive HM is superconducting we conclude
from the above that the repulsive HM is also superconducting.

Further light on the mechanism of pairing in the HM
comes from an independent
study of the Hubbard model based on the Roth decoupling
procedure~\cite{22}.
This study shows pairing of $d_{x^2-y^2}$
symmetry, and exhibits a clear correlation between pairing and
nearest-neighbor spin correlation.  Therefore the source of the
pairing is probably a nearest neighbor antiferromagnetic exchange
interaction, generated by higher order
diagrams~\cite{23} in U, starting
at $U^2$. \\

In conclusion, pairing correlations in $d_{x^2-y^2}$
symmetry, but no other symmetry,
are found for the clusters and parameters studied
in this paper. We presented two different ways of extrapolating
our data to the thermodynamic limit.
The $tt'$ HM phase diagram thus includes
the Mott transition, marginal Fermi Liquid behavior, and
superconductivity - all the key features
present qualitatively in real cuprates.
The pairing interaction found in the parameter space
explored so far supports a $T_c$ of order 40 K.
Full understanding of high temperature superconductivity,
including the highest $T_c$'s,
probably requires consideration
of additional interactions, such as superexchange from the
oxygen bands and mediation by phonons. \\

The Monte Carlo simulation was carried out on the IBM SP1.
Typically the simulation at a single parameter occupied 4 processors for
80 hours.

Acknowledgement

   Part of this work was supported by the Bayerische Forschungsverbund
FORSUPRA.
We are grateful to Joefon Jann for
assistance with parallelization, and to K.A. M\"uller, D.J. Scalapino,
and C.C. Tsuei for valuable discussions.

\newpage

\section*{Figure Captions}
\begin{description}
\item[Fig. 1]
Fermi surface of noninteracting $tt'$-Hubbard model
$t'= 0.3 t$, $E_F = -1.2t$, hole doping
= 0.27.  Solid circles indicate saddle points.
\item[Fig. 2a]
Superconducting vertex correlation function $\chi(R)$
(Eq. (3)), plotted versus distance $R$, $12\times 12$ lattice,
106 electrons (doping 0.264), $\Theta = 8/t$,
$t'= 0.286t$, $U = 2t$. Insert shows correlation function
on expanded scale, horizontal dashed line is average plateau value.
Error bars are less than the width of the points.
\item[Fig. 2b]
QMC calculation of plateau vs. U, for $8\times 8$ lattice,
50 electrons (doping 0.22),
$\Theta  = 8/t$, $t'=0.22$
\item[Fig. 3]
Filled points QMC calculation of
plateau value $\chi^{pl}$ of superconducting vertex correlation
function
(Eq. (3)) versus $t'$, for
$10\times 10$ lattice, 74 electrons (doping 0.26), $\Theta = 8/t$,
$U = 2t$, error bar is average; open points fbcs calculation (Eq. (4))
with J=0.055t and cutoff $omega_c = 0.2 t$.
\item[Fig. 4]
Superconducting vertex correlation $\chi^{pl}$ in the onsite-s channel
for attractive $U=-0.3$ versus superconducting vertex correlation $\chi^{pl}$
in the
$d_{x^2-y^3}$-channel for the repulsive Hubbard model ($U=2$). Filling
and $t'$ are the same for both values of U and are indicated in table I.
\end{description}

\newpage
\begin{table}[h]
\begin{tabular}{ccccc}
$N_L$ & $N_e$ & $\omega_c$ & $\chi^{pl}$ & $J$ \\
 36   & 26    & 0.3 t      & 1.377E-03   & 0.12 t \\
 64   & 50    & 0.25 t     & 0.648E-03   & 0.15 t \\
100   & 82    & 0.25 t     & 0.491E-03   & 0.15 t \\
144   & 122   & 0.25 t     & 0.332E-03   & 0.15 t
\end{tabular}
\label{tab1}
\end{table}

\begin{description}
\item[Table I] Scaling $t'=0.22 t$, $U=2t$
\end{description}


\begin{thebibliography}{99}
\bibitem[1]{1} Monthoux and D. Pines, Phys. Rev. Lett., {\bf 69} 961 (1992);
ibid., Phys.Rev. B {\bf 49}, 4261 (1994).
\bibitem[2]{2} Monthoux and D.J. Scalapino, Phys. Rev. Lett. {\bf 72},
1874 (1994); D.J. Scalapino, Physica C {\bf 185-189}, 104 (1991).
\bibitem[3]{3} C. Tsuei et al., Phys. Rev. Lett. {\bf 73},
593 (1994); J. R. Kirtley et al., Nature
{\bf 373}, 225 (1995).
\bibitem[4]{4} L. N. Vu et al., Appl. Phys. Lett.
{\bf 63}, 1693 (1993).
\bibitem[5]{5} J.E. Sonier et al., Phys. Rev. Lett. {\bf 72}, 744 (1993);
M. Takigawa, J.L. Smith and W.L. Hults, Phys. Rev. {\bf B44}, 7764 (1991);
M. Takigawa and D.B. Mitzi "NMR studies of spin excitations in
superconducting $\rm Bi_2 Sr_2 Ca Cu_2 O_{2+\delta}$
single crystals", submitted to Phys. Rev. Lett..
\bibitem[6]{6} M. Imada, Physica C {\bf 185-189}, 1447 (1991),
N. Furukawa and M. Imada, J. Phys. Soc. Japan {\bf 61},
3331 (1992).
\bibitem[7]{7} W. von der Linden, Phys. Repts. {\bf 220}, No.2,3 (1992).
\bibitem[8]{8} A. Moreo, Phys. Rev. {\bf B45}, 5059 (1992).
\bibitem[9]{9} S.R. White et al., Phys. Rev. {\bf B39}, 839 (1989).
\bibitem[10]{10} J. Labbe and J. Bok, Europhys. Lett. {\bf 3}, 1225 (1987).
\bibitem[11]{11} R.S. Markiewicz, Physica C {\bf 217}, 381 (1993),
and references therein.
\bibitem[12]{12} D. M. Newns et al., Comments Condens. Matter Phys.
{\bf 15},  273-302 (1992);
P.C. Pattnaik et al., Phys. Rev. {\bf B45}, 5714 (1992);
D.M. Newns et al., Phys. Rev. Lett. {\bf 73}, 1695 (1994);
D.M. Newns et al., 'van Hove Scenario for d-wave superconductivity
in Cuprates', submitted to Physical Review B.
\bibitem[13]{13} D.M. King et al., Phys. Rev. Lett. {\bf 73}, 3298 (1994);
D.S. Dessau et al., Phys. Rev. Lett. {\bf 71}, 2781 (1993);
D.M. King et al., Phys. Rev. Lett. {\bf 70}, 3159 (1993).
\bibitem[14]{14} K. Gofron et al.,
Phys. Rev. Lett. {\bf 73}, 3302 (1994);
A.A. Abrikosov, J.C. Campuzano and K. Gofron, Physica C
{\bf 73}, 214 (1993).
\bibitem[15]{15} J.E. Hirsch, Phys. Rev. {\bf B31}, 4403 (1985).
\bibitem[16]{16} C.N. Yang, Rev. Mod. Phys. {\bf 34}, 694 (1962).
\bibitem[17]{17} W. Fettes et al., "Density of States in the $t'$ Hubbard
Model, Exact diagonalization Results", preprint.
\bibitem[18]{18} J.W. Serene and D.W. Hess,
"Recent Progress in Many-Body Theories, vol. 3,"
edited by T.L. Ainsworth, C.E. Campbell, B.E. Clements, and
E. Krotscheck (Plenum, New York, 1992), p. 469.
\bibitem[19]{19} H.-B. Schuttler and A.J. Fedro, Phys. Rev. {\bf B45},
7588 (1992).
\bibitem[20]{20} J. Wheatley and T. Xiang, Solid State Commun. {\bf 88},
593 (1993).
\bibitem[21]{21} J. Rossat-Mignod et al., Physica B {\bf 186-188}, 1 (1993);
T.E. Mason et al., Phys. Rev. Lett. {\bf 71}, 919 (1993);
S-W. Cheong et al., Phys. Rev. Lett. {\bf 67}, 1791 (1993).
\bibitem[22]{22} J. Beenen and D.M. Edwards,
Proc. Int. Euroconf. on Magnetic Correlations, Metal-Insulator
Transitions and Superconductivity in Novel Materials (Wurzburg 1994);
J. Low. Temp. Phys. (to be published).
\bibitem[23]{23} A. Kampf and J.R. Schrieffer, Phys. Rev. {\bf 41},
6399 (1990).
\bibitem[24]{24} E.Y. Loh(Jr), J.E. Gubernatis, "Electronic Phase Transitions"
edited by W. Hanke, Yu. V. Kopaev (North-Holland, Amsterdam 1989)
\bibitem[25]{25} H. de Raedt, M. Frick, Phys. Rep. {\bf 231} 107 (1992)
\bibitem[26]{26} S. Sorella, Baroni S.,
Car R., Parinello M., Europhysics Letter {\bf 8} 663 (1989)
\bibitem[27]{preprint} T. Husslein, W. Fettes, I. Morgenstern to be published
\end{thebibliography}
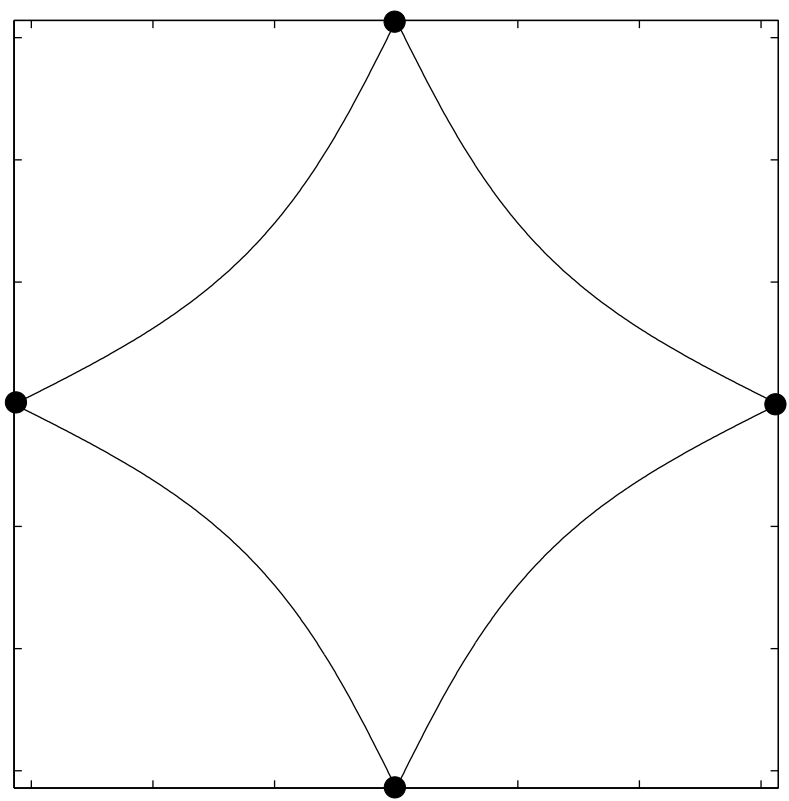
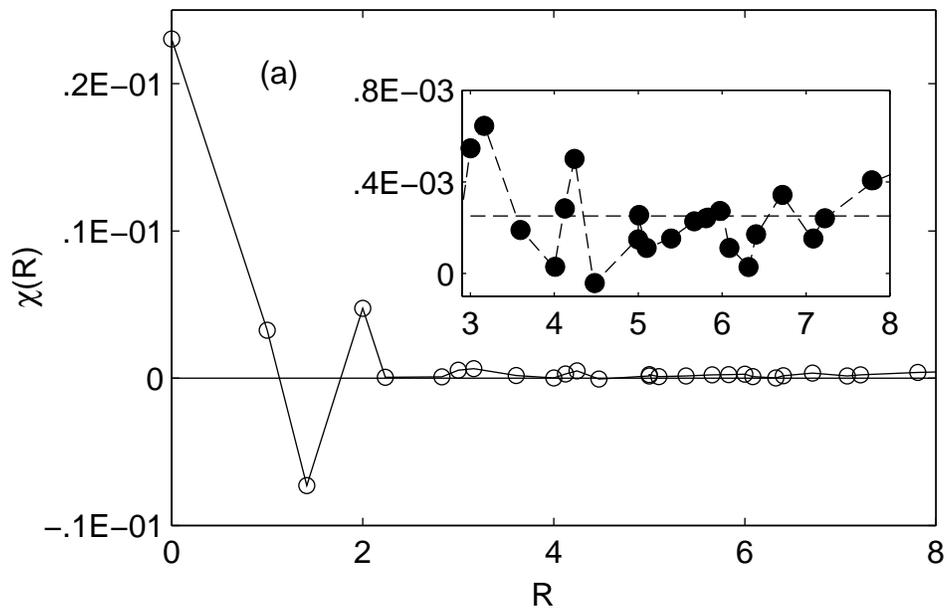
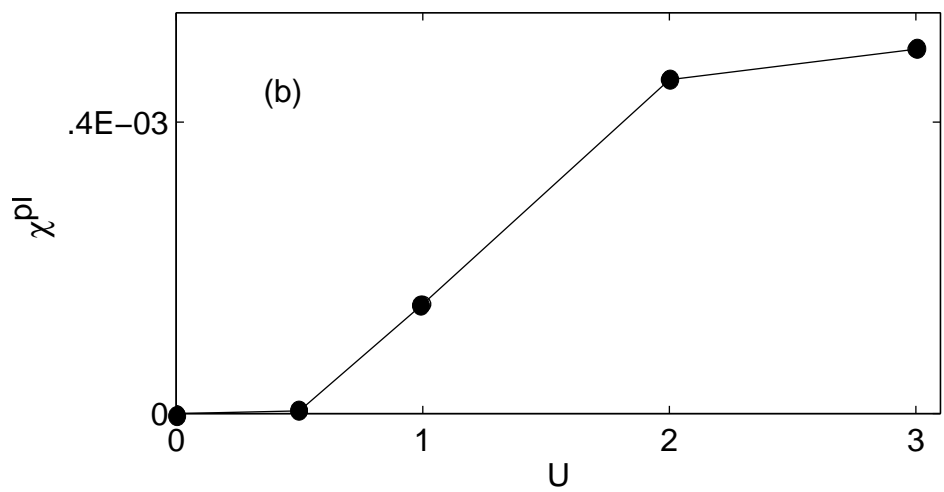
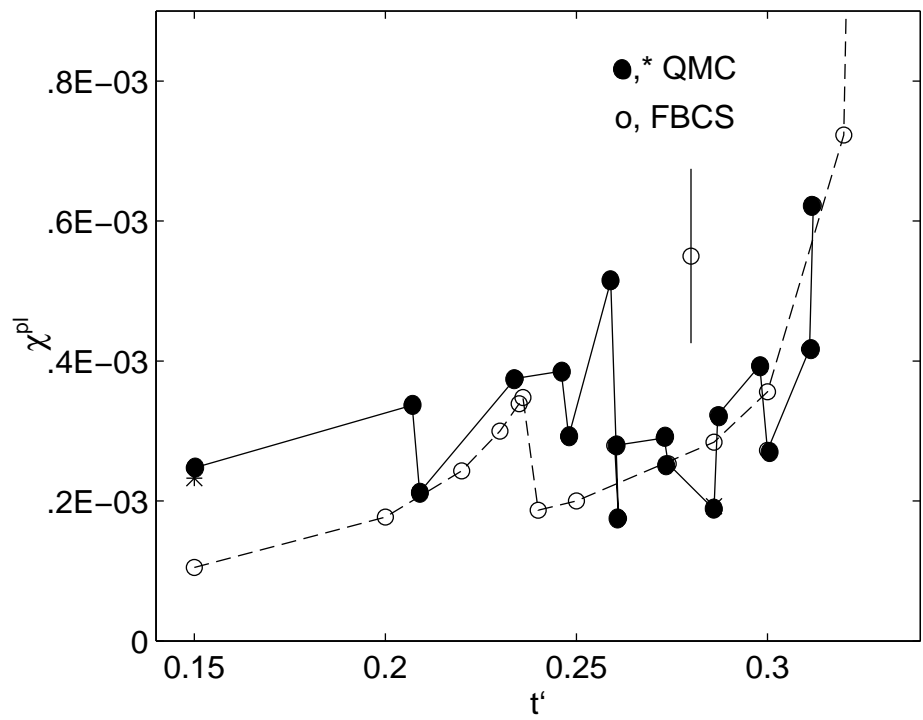
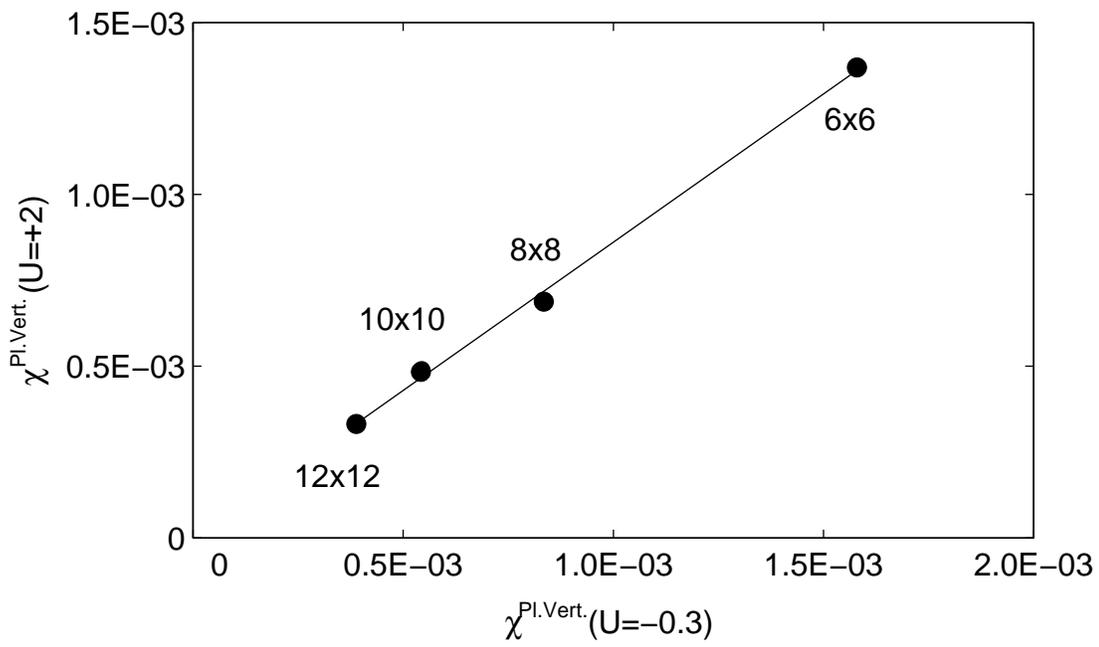
\end{document}